\documentclass[preprint,12pt]{elsarticle}

 \usepackage{graphics}
 \usepackage{graphicx}
\usepackage{epsfig}
\usepackage{CJK}
\usepackage{amssymb}
\usepackage{amsthm}
\usepackage{amsmath}
\usepackage{mathrsfs}
\usepackage{array,graphicx,subfig}
\usepackage{algorithmic}
\usepackage{algorithm}
\usepackage{booktabs}
\usepackage{longtable}
\usepackage{tabularx}

\newtheorem{definition}{Definition}[section]
\newtheorem{coro}{Corollary}[section]

\theoremstyle{definition}

\newtheorem{prop}{Proposition}[section]
\newtheorem{exam}{Example}[section]
\newtheorem{thero}{Theorem}[section]
\newtheorem{lemma}{Lemma}[section]

\begin{document}

\begin{frontmatter}



\title{A new type of judgement theorems for attribute characters in information system}
\author[a]{Anhui Tan\corref{cor1}} \cortext[cor1]{Corresponding author.} \ead{shujujiegouwang@126.com}\vskip 2mm
\author[b]{Jinjin Li} \ead{jinjinli@fjzs.edu.cn}\vskip 2mm
\author[b,c]{Guoping Lin}\ead{guoplin@163.com}\vskip 2mm
\address[a]{School of Mathematical Sciences, Xiamen University, Xiamen 361005, P. R. China}\vskip 3mm
\address[b]{Department of Mathematics and Information Science, Zhangzhou Normal University, Zhangzhou, 363000, Fujian, China} \vskip 3mm
\address[c]{School of computer and Information Technology, Shanxi University, Taiyuan, 030006, Shanxi, China} \vskip 3mm

\begin{abstract}
The research of attribute characters in information system which contains core, necessary, unnecessary is a basic and important issue in attribute reduct. Many methods for the judgement of attribute characters are based on the relationship between the objects and attributes. In this paper, a new type of judgement theorems which are absolutely based on the relationship among attributes is proposed for the judgement of attribute characters. The method is through comparing the two new attribute sets~$E(a)$~and~$N(a)$~with respect to the designated attribute~$a$~which is proposed in this paper. We conclude that which type of the attribute~$a$~belongs to is determined by the relationship between~$E(a)$~and~$N(a)$~in essence. Secondly, more concise and clear results are given about the judgment of the attribute characters through analyzing the properties of refinement and precise-refinement between~$E(a)$~and~$N(a)$~in topology. In addition, the relationship among attributes are discussed which is useful for constructing a reduct in the last section of this paper. In the last, we propose a reduct algorithm based on~$E(a)$, and this algorithm is an extended application of the analysis of attribute characters above.
\end{abstract}


\end{frontmatter}


\section{Introduction}
Rough set theory originally proposed by Pawlak $[1][2]$, provides an effective method to deal with imprecision and vague in information system. Because of voluminous data that always generates in information system, multi-granular Rough set is very important as a basic Processing method, and has very widely applications in process control, conflict analysis, data mining, data fusion technique $[18,19,20,21]$. In recent years, many topics have been widely investigated with rough sets, for example, topology, graph, algebra, lattices, fuzzy set, and so on $[24-33]$.

Information system is an extension of rough set theory, which is denoted as a pair $(U,A)$, where~$U$~is a nonempty finite set of objects called the domain set and~$A=\{a_{1},a_{2}, ... ,a_{n}\}$~is a nonempty finite set of attributes such that~$a:U\rightarrow V_a$~for any~$a\in A,$~i.e.,~$a{x}\in V_a, x\in U,$~where~$V_{a}$~is called the domain of attribute~$a$. Each attribute based on an equivalence relation generates a partition on the domain~$U$, and the objects belong to the same partition have the same value in $V_{a}$, so the set of attributes~$A$~generates more than one partitions.

Moreover, approximation space theory is proposed to approximate the objects that are not precise. In approximation space theory, a pair of approximation operators~$R_{*},R_{*}$ are introduced which are widely used.

Because of the importance of the attributes, attribute reduction is a basic topic in rough sets theory and information system, and it is a NP-hard problem.

A beautiful and efficient method was introduced by Skowron and Rauszer $[6]$ which is the  discernibility matrix of the information system $(U,A)$, where $D_{A}=\{d_{A}(x,y)|x,y\in U\}$, $d_{A}(x,y)=\{a\in A|V_{a}(x)\neq V_{a}(y)\}$. Otherwise, the boolean discernibility function is proposed to compute the reduct. Many scholars devote to study this issues $[11-14],[22,23]$. Especially, many papers studied the attribute reduct theory with the discernibility matrix$[15-21]$.

In $[3]$, Yao and Zhao discussed the reduct by discernibility matrix simplification. The thought is through deleting the attributes in every ~$d_{A}(x,y)$, and in the end, there is only one attribute in every $d_{A}(x,y)$, then the remaining attributes constitute a reduct. Otherwise, another reduct construction algorithm based on matrix simplification is proposed in [3]. The computing process of this algorithm is just like the operation of the boolean discernibility function. That is, delete the redundant elements in the discernibility matrix through absorption operation, then there exists no inclusion relationship between any two elements of the discernibility matrix. Then pick a $d_{A}(x,y)$ in the discernibility matrix, and an attribute ~$a\in d_{A}(x,y)$, then delete~$(d_{A}(x,y)-\{a\})$~for any other elements of the discernibility matrix. Then return to repeat the algorithm until there is only one attribute in every element of the discernibility matrix. Then the algorithm ends and a reduct is constructed.

Otherwise, In [4], Zhang and Qiu depart the attributes into three types based on the relationship with attribute reduct which contain core, relative necessary and unnecessary attributes. Many beautiful theorems are given to describe these three type attributes. However, they are all not substantive description for attribute character.

In this paper, we propose an attribute subset~$E(a)$~with respect to the attribute~$a$. We conclude that whether the attribute~$a$~is relative necessary or unnecessary is determined by~$E(a)$~virtually. Secondly, refinement and precise-refinement in topology are used to study the properties of attributes. Moreover, combined with the row-wise simplification reduct construction algorithm in [3], we construct a algorithm based on~$E(a)$.

The remainder of this paper is organized as follows. Some basic concepts in rough set theory are reviewed in Section 2. The three types of attributes are investigated in Section 3, and more substantial descriptions to the three types are given in this section. In Section 4, With the tool of topology, more clear and easier conclusions are given about the three types of attributes.

Because of the reduct  based on the relationship between the attributes virtually, in section 5, the relationship between the attributes are discussed. In the last of this paper, we give a reduct algorithm based on~$E(a)$, and analysis it's efficiency through an example in [3].\\

\section {Preliminaries}
In this section, we review some basic concepts, which includes
Pawlak's rough set, information system, and
covering-based rough set. (see [1, 4, 5, 6, 16, 25, 27, 28])
\subsection{Pawlak's rough set}
Rough set theory, proposed by Pawlak , provides a useful approach to deal with dramatic data. In classical  Rough set theory, the data is divided into some equivalency classes.
\begin{definition}
(Approximation space) Let~$U$~be a nonempty and finite set and~$R$~be an equivalence relation on~$U$, i.e., ~$R$~is reflexive, symmetric and transitive. The ordered pair~$(U,R)$~is called an approximation space.
\end{definition}
So in an approximation space~$(U,R)$,~$R$~generates a partition~$U/R=\{X_{1},X_{2},...,X_{m}\}$~on~$U$,
where~$X_{1},X_{2},...,X_{m}$~are the equivalence classes generated by the equivalence relation~$R$.

In rough set, a pair of approximation operators are used to describe the objects. In the following definition, a pair of approximation operators are introduced which are widely used.
\begin{definition}
(Approximation operator) Let~$U$~be a nonempty and finite set and~$R$~be an equivalence relation on~$U$, a pair of approximation operator~$R_{*}(X)=\{x\in U|RN\{x\}\subseteq X\},$~~$R_{*}(X)=\{x\in U|RN\{x\}\cap X\neq \emptyset\}.$~Where~$RN(x)=\{y\in U|xRy\}$. They are called the lower and upper approximation operators with respect to~$R$, respectively.
\end{definition}
\begin{definition}
(R-precise and R-rough set) Let~$R$~be an equivalence relation on~$U$. For all~$X\subseteq U,$~if ~$R_{*}(X)=R^{*}(X),$~then~$X$~is a~$R-$precise set; otherwise, we say~$X$~is a~$R-$rough set.
\end{definition}
\subsection{Information system}
The notion of information systems is an extension of rough set, and provides a convenient tool for the representation of objects in terms of their attribute values.
An information system $\textbf{IS}$ is a pair $(U,A)$, where~$U$~is a nonempty finite set
of objects called the domain set and $A=\{a_{1},a_{2}, ... ,a_{n}\}$ is a nonempty finite set of attributes such that~$a:U\rightarrow V_a$~for any~$a\in A,$~i.e.,~$a{x}\in V_a, x\in U,$~where~$V_a$~is called the domain of attribute~$a$.

Each nonempty subset~$B\subseteq A$~in an $\textbf{IS}$ determines an indiscernibility relation as follows:
~$R_{B}=\{(x,y)\in U*U:a(x)=a(y),\forall a\in B\}.$

Since~$R_{B}$~is an equivalence relation on~$U$, it forms a partition~$U/R_{B}={[x]_B:x\in U},$ where~$[x]_B$~ is the
equivalence class determined by~$x$~with respect to~$B$, i.e.,~$[x]_B =\{y\in U:(x,y)\in R_{B}\}.$~$U/R_{B}$~reflects the basic granules of knowledge w.r.t. $B$ in the $\textbf{IS}$.
\begin{definition}
(consistent)An information system $(U,A)$,~$B\subseteq A$, $B$~is a consistent set iff~$U/R_{B}=U/R_{A}$.
\end{definition}
\begin{definition}
(reduct)An information system $(U,A)$,~$B\subseteq A$, $B$~is a reduct iff~$B$~is a consistent set and for any ~$C\subset B$, $U/R_{C}\neq U/R_{A}$.
\end{definition}
Let we denote a set~$B$~can not be reducted iff for all ~$C\subset B$, $U/R_{C}\neq U/R_{B}$.
\begin{definition}
In an information system~$(U,A)$, We divide the the attribute set~$A$~into three types as follows:

(1)(core) For~$a\in A$, $a$~is a core attribute iff for all reduct set~$B$, $B\subseteq A$, $a\in B$.

(2)(unnecessary) An information system $(U,A)$, for~$a\in A$, $a$~is an unnecessary attribute iff for all reduct ~$B$, $B\subseteq A$, but $a\notin B$.

(3)(relative necessary) An information system $(U,A)$, for~$a\in A$, $a$~is a relative necessary attribute iff~$a$~is not a core attribute and there exists a reduct~$B$, $B\subseteq A$, and $a\in B$.
\end{definition}
In an information system~$(U,A)$, we denote~$(U,B_{i}),i\in \tau$ for all the reduct subsets of~$A$,~$\tau$~is an index set. So we denote the core attribute set as~$C=\cap_{i\in \tau}B_{i}$, the set of all the relative necessary attributes as~$RN=\cup_{i\in \tau}B_{i}-\cap_{i\in \tau}B_{i}$, and the set of all the unnecessary attributes as~$UN=A-\cup_{i\in \tau}B_{i}$.

 Every attribute can generates an equivalent relation and generates a partition, so in rough set theory, the study of attribute is very important, and discernibility matrix leads an important role to study the feature of attributes.
\begin{definition}
(discernibility matrix)An information system~$(U,A)$,~$|U|=n$, for~$x,y\in U$,~$d_{A}(x,y)=\{a\in A|V_{a}(x)\neq V_{a}(y)\}$. we say~$d_{A}(x,y)$~is a partition discernibility set of~$x,y$.
We denote~$D_{A}=\{d_{A}(x,y)|x,y\in U\}$~for the discernibility matrix of the information system~$(U,A)$.
\end{definition}
\begin{definition}
(discernibility function)The discernibility function of a discernibility matrix is defined by:
$f(D_{A})=\bigwedge\{\bigvee d\{x,y\}|d\{x,y\}\in D_{A},x,y\in U\}$.
\end{definition}

\subsection{Covering-based rough set}
\begin{definition}
 Let~$U$~be the domain of discourse and~$C$~a family of subsets of~$U$. If none of the subsets in
 ~$C$~is empty, and~$\bigcup C = U$,~$C$~is called a covering of~$U$.
 \end{definition}
Since it is clear that a partition is definitely a covering, the concept of coverings is an extension of the concept of partitions.
\begin{definition}
(Covering approximation space). Let U be a non-empty set and~$\mathcal{C}$~ be a covering of U. We call the ordered pair~$U,\mathcal{C}$~a covering approximation space.
\end{definition}
\begin{definition}
(Minimal description). Let~$(U,C)$~be a covering approximation space. If~$x\in U$, the minimal description of~$x$~ is defined as~$Md(x)=\{K\in C|x\in K,\wedge (\forall S\in C \wedge x\in S\subseteq K\Rightarrow K=S)\}$.
\end{definition}

Paper $[1,16]$ introduced a new definition for binary relation-based rough sets. The concept for this definition is the neighborhood of a point.In the following we introduce the neighborhood concept into covering-based rough
set.
\begin{definition}
(Neighborhood) Let $U$ be a domain of objects, $C$ a covering of $U$. For any~$x\in U$, we define the neighborhood of
$x$ as follows:
~$Neighbore(x)=\{\cap K|x\in K\in C\}$.
\end{definition}
The concept of~$Neighbore(x)$~is proposed to study the covering-based rough set, and let us denote a new concept ~$N(x)$ that just like the notion of~$Neighbore(x)$, but they are different. In an information system~$(U,A)$, the partition discernibility sets $d_{A}(x,y)$ in a discernibility matrix are always used, so we denote~$N(a)=\{d_{A}(x,y)|a\in d_{A}(x,y),x,y\in U\}$ for the attribute~$a$.

In the following, we will study the covering-based rough set of attribute with
~$Neiborhood(x)$~and~$N(x)$.

Let~$(U,C)$~be a covering approximation space and $X\subseteq U$.
\begin{definition}
(Lower approximation) The covering lower approximation operation~$R_{*}:P(U)\Rightarrow P(U)$~is defined as ~$R_{*}(X)=\bigcup_{K\in C\wedge K\subseteq X}K$$[28]$.
\end{definition}
\begin{definition}
 (Upper approximation) Let $C$ be a covering of the domain $U$. The covering upper approximation operation~$R^{*}$, is defined as: $X\subseteq U, R^{*}(X)= \bigcup_{K\in C\wedge K\cap X\neq \emptyset}K$.
\end{definition}
\begin{prop}
Let $C$ be a covering of $U$, the followings are equal.
(1)~$\{x\}\in C$.
(2)~$Md(x)=\{\{x\}\}$.
(3)~$R_{*}(x)=\{x\}$.
(4)~$Md(x)=\{R_{*}(x)\}$.
\end{prop}
\textbf{Proof.} $(1)\Leftrightarrow (2)\Leftrightarrow (3))$ is easy to proof. So we only proof~$(1)\Leftrightarrow (4)$.\\
$(4)\Rightarrow (1)$: since $C$ be a covering of $U$, then there exists $x\in K\in C$, thus
~$Md(x)\neq\emptyset$, and~$Md(x)=\{R_{*}(x)\}$, so $R_{*}(x)\neq\emptyset$, thus $R_{*}(x)=\{x\}$, so $\{x\}\in C$.  \\
$(1)\Rightarrow (4)$: if $\{x\}\in C$, then $R_{*}(x)=\{x\}$, then ~$Md(x)=\{R_{*}(x)\}$.
\section{The analysis of attribute characters in information system}
 An information system~$(U,A)$,~$|U|=n$,~for~$x,y\in U$,~$d_{A}(x,y)=\{a\in A|V_{a}(x)\neq V_{a}(y)\}$.
the discernibility matrix of the information system~$(U,A)$ is~$D_{A}=\{d_{A}(x,y)|x,y\in U\}$. Let the covering ~$\mathcal{C}=\{d_{A}(x,y)\}$,~for all~$x,y\in U$, then~$\mathcal{C}$~is a covering of~$A$.
We denote~$C_{D}=\{d_{A}(x,y),x,y\in U\}$~the covering-based rough sets of attribute.

Otherwise, for~$B\subseteq A$, we denote~$R_{*}(B)=\{d(x,y)|d(x,y)\in D_{A},d(x,y)\subseteq B\}$,
~$R^{*}(B)=\{d(x,y)|d(x,y)\in D_{A},d(x,y)\cap B\neq\emptyset\}$.~So~$R_{*}$~and~$R^{*}$~
in attribute covering are corresponding to those concepts in classic rough set.
\begin{thero}
An information system $(U,A)$,~$B\subseteq A$, $B$~is a consistent set iff~$\forall x,y\in U,d_{A}(x,y)\neq\emptyset,$~then~$B\cap d_{A}(x,y)\neq\emptyset$.
\end{thero}
\textbf{proof.} Sufficiency. We need to proof~$U/R_B=U/R_A$. Because~$[x]_B =\{y\in U:(x,y)\in R_B\}$, if
 ~$B\cap d_{A}(x,y)\neq\emptyset$, then for~$y\in [x]_A$, then for any ~$a\in A$,~$y\in [x]_a$, and~$B\subseteq A$, then~$y\in [x]_B$. For~$y\notin [x]_A$, there exists~$a\in A$,~$y\notin [x]_a$, then~$a\in d_{A}(x,y)$, and~$B\cap d_{A}(x,y)\neq\emptyset$, then there exists~$b\in B$,~$y\notin [x]_b$, then~$y\notin [x]_B$. So~$U/R_B=U/R_A$.

 Necessity. If~$(U,B)$~is a consistent information system, then~$U/R_B=U/R_A$. For~$y\notin [x]_A$, then For~$y\notin [x]_B$, there exists~$b\in B$,~$y\notin [x]_b$, then ~$b\in d_{A}(x,y)$, then
 ~$B\cap d_{A}(x,y)\neq\emptyset$.~$\Box$
\begin{thero}
An information system $(U,A)$,~$B\subseteq A$, $B$~is a reduct iff~$\forall x,y\in U,d_{A}(x,y)\neq\emptyset,$~then~$B\cap d_{A}(x,y)\neq\emptyset$,~and for any~$a\in B$,~$B-\{a\}$ does not holds.
\end{thero}
\textbf{Proof.}It is obvious from Theorem 3.1. $\Box$
\begin{coro}
An information system $(U,A)$,~$B\subseteq A$, $B$~is a reduct set iff ~$\forall a\in A, \forall K\in N\{a\}\in U$, then~$B\cap K\neq\emptyset$.
\end{coro}
\textbf{Proof.}It is obvious from Theorem 3.1.~$\Box$
\begin{coro}
An information system $(U,A)$, ~$B\subseteq A$, $B$~is a reduct set, if for ~$a\in A$,~for every~$K\in N(a)$, ~$(B-\{a\})\cap K\neq\emptyset$, then~$a\notin B$.
\end{coro}
\textbf{Proof.}If for every~$K\in N(a)$,~$(B-\{a\})\cap K\neq\emptyset$, then~$R_{B-\{a\}}=R_{B}$, then~$B$~can be reducted.

If~$\{a\}$~is an element of~$C_{D}$, we assume~$d_{A}(x,y)=\{a\}$. then~$y\in [x]_(A-\{a\})$, and~$y\notin [x]_A$. then~$U/R_(A-\{a\})\neq U/R_A$, so for any reduction~$(U,B)$,~$B\nsubseteq (A-\{a\})$, so~$a\in B$.

 If ~$a\in A$~is a core attribute, then for any reduction~$(U,B)$,~$a\in B$. If~$\{a\}$~is an element of~$C_{D}$, then for~$x,y$, if~$a\in d_{A}(x,y)$, then there is another element~$b\in A$,~$b\in d_{A}(x,y)$, then we can proof there is a reduct set~$C$,~$b\in C$, and~$a\notin C$. $\Box$
\begin{coro}
~$C_{D}=\{d_{A}(x,y),x,y\in U\}$~the covering-based rough sets of attribute,~$a\in A$~is a core attribute
iff~$Md(x)=\{\{x\}\}$.
\end{coro}
 \begin{coro}
~$C_{D}=\{d_{A}(x,y),x,y\in U\}$~the covering-based rough sets of attribute,~$a\in A$~is a core attribute
iff~$R_{*}(x)=\{x\}$.
\end{coro}
 \begin{coro}
~$C_{D}=\{d_{A}(x,y),x,y\in U\}$~the covering-based rough sets of attribute,~$a\in A$~is a core attribute
iff~$Md(x)=\{R_{*}(x)\}$.
\end{coro}
\begin{thero}
$[6]$ In an information system~$(U,A)$, attribute~$a\in A$~is a core iff there exists some~$d\{x,y\}, x,y\in U$,~$d\{x,y\}\in D_{A}$, such that~$d\{x,y\}=\{a\}$ iff~$R_{A-\{a\}}\neq R_{A}$.
\end{thero}
\begin{thero}
$[4]$ In an information system~$(U,A)$, we have  the conclusions as follows:\\
 (1)~$a\in A$~is a relative necessary attribute iff ~$R_{A-\{a\}}= R_{A}$, and~$ \cup\{R_{B-\{a\}}\nsubseteq R_{a}$~,where~$R_{B}\subseteq R_{A},B\subseteq A\}$.\\
 (2)~$a\in A$~is a unnecessary attribute iff ~$\cup\{R_{B-\{a\}}\}\subseteq R_{a}$, where~$R_{B}\subseteq R_{A},B\subseteq A$.
\end{thero}

Theorem 3.4 describes the character of attributes, but it's just a kind of representation in definition, and lack of real substance in some degree. What the theorem wants to express is as follows:

(a)~$R_{A-\{a\}}= R_{A}$~~$\Leftrightarrow$~ when deleting ~$a$~from ~$A$, the partition doesn't change~$\Leftrightarrow$~$a$~is not a core.

(b)~$R_{B}\subseteq R_{A}$$\Leftrightarrow$$B$~is a consistent set of $A$.

(c)~$ \cup R_{B-\{a\}}\nsubseteq R_{a}$, where~$R_{B}\subseteq R_{A}$~$\Leftrightarrow$~there exists a set~$B_{0}$, satisfies~$R_{B_{0}-\{a\}}\nsubseteq R_{a}$, where~$R_{B_{0}}\subseteq R_{A}\}$
$\Leftrightarrow$there exists a reduct~$B_{0}$~satisfies~$R_{B_{0}-\{a\}}\nsubseteq R_{a}$$\Leftrightarrow$
$a\in B_{0}$$\Leftrightarrow$$a$~belongs to some reduct.

(d) $R_{A-\{a\}}= R_{A}$,~$ \cup R_{B-\{a\}}\nsubseteq R_{a}$, where~$R_{B}\subseteq R_{A}$$\Leftrightarrow$~
$a$~is not a core, and~$a$~belongs to some reduct$\Leftrightarrow$$a$~is a relative necessary attribute.

(e)~$\cup\{R_{B-\{a\}}\}\subseteq R_{a}$, where~$R_{B}\subseteq R_{A},B\subseteq A$$\Leftrightarrow$~
for every reduct ~$B$, the partition generated by~$a$~is coarser than~$B-\{a\}$$\Leftrightarrow$
for every reduct~$B$,~$a\notin B$$\Leftrightarrow$$a$~is a unnecessary attribute.

From the explanation above, we can see that the therom 4.2 doesn't give more essential information. For example,
In Theorem 3.4(2), it means for all reduct~$B$, when deleting~$a$~from~$B$, the partition doesn't become finer, then ~$a$~is unnecessary. In the following, we will give some equivalence conditions to describe the feature of attributes with more substantial meaning and value.

We denote~$E(a)=\{d\{x,y\}|a\notin d\{x,y\}\subseteq \cup N(a), d\{x,y\}\in D_{A}\}$. We conclude that which type of attribute~$a$~belongs to is determined by the relation between~$N(a)$~and~$E(a)$.
\begin{lemma}
~$\cup E(a)=R_{*}(\cup N(a)-\{a\})$, and~$E(a)=\{d\{x,y\}| d\{x,y\}\subseteq (\cup N(a)-\{a\}), d\{x,y\}\in D_{A}\}$.
\end{lemma}
Proof. It's obvious. ~$\Box$
\begin{thero}
~$C_{D}=\{d_{A}(x,y),x,y\in U\}$~the covering-based rough sets of attribute,~$a\in A$~is a unnecessary attribute iff for any set~$C\subseteq A$,~$R_{C}\subseteq R_{\cup E(a)}$,$\Rightarrow$$R_{C}\subseteq R_{a}$.
\end{thero}
\textbf{Proof.}
Sufficiency.~$R_{*}\{\cup N(a)-\{a\}\}=\cup \{M\in C_{D},M\subseteq N(a)-\{a\}\}$,~$R^{*}C=\cup \{N\in C_{D},N\cap C\neq \emptyset\}$.

For~$C\subseteq \cup E(a)$,~$R_{C}\subseteq R_{\cup E(a)/}$,~$R_{C}\subseteq R_{\cup E(a)/}\Rightarrow C\cap \{M\in C_{D},M\subseteq N(a)-\{a\}\}\neq \emptyset$, then for any reduct~$B$, there exists $C\subseteq \cup \{M\in C_{D},M\subseteq N(a)-\{a\}\}$,~$C\subseteq B$,~$C\cap \{M\in C_{D},M\subseteq N(a)-\{a\}\}\neq \emptyset$. If
~$C\cap \{M\in C_{D},M\subseteq N(a)\}\neq \emptyset$, then~$B\cap \{M\in C_{D},M\subseteq N(a)\}\neq \emptyset$. Then ~$a\notin B$.

Necessity. If~$a\in A$~ is a unnecessary attribute, then for any reduction~$(U,B)$,~$a\notin B$,
because~$\forall K\in N\{a\},$ ~$B\cap K\neq \emptyset$, there exists
~$C\subseteq \cup \{M\in C_{D},M\subseteq N(a)-\{a\}\}$,~$C\subseteq B$, if there is a set~$K\in N\{a\}$,
~$C\cap K=\emptyset$, and we can proof~$B\cap K=C\cap K $, then~$B\cap K=\emptyset$, then ~$a\in B$.Because~$a$~is a unnecessary attribute, then it's a contradiction. $\Box$
\begin{thero}
~$C_{D}=\{d_{A}(x,y),x,y\in U\}$~the covering-based rough sets of attribute,~$a\in A$ is a unnecessary attribute iff for any~$K\in E(a)$, and any~$C\subseteq A$, satisfying~$C\cap K \neq \emptyset$, then for any $F\in N(a)$, satisfying~$C\cap F \neq \emptyset$.
\end{thero}
\textbf{Proof.} Sufficiency.
Because for reduct ~$B$, for any set~$K\in E(a)$, satisfies~$B\cap K \neq \emptyset$, let $C=B\cap \cup K$,~$K\in E(a)$, then for any $F\in N(a)$,~$C\cap K \neq \emptyset$, then ~$B\cap F \neq \emptyset$, then ~$A\notin B$.

Necessity.
If there exists a set~$C$,~$C\subseteq A$, for any set~$K\in E(a)$,~$C\cap K \neq \emptyset$, but there exists a set ~$F_{0}$, $F_{0}\in N(a)$,~$C\cap F_{0}=\emptyset$.
 When  constructing any reduct~$B$, we must pick the attributes in all the~$d\{x,y\}$,~$d\{x,y\}\in D_{A}$. Assuming we have pick an attribute set $C$~in every~$F$, $F\in E(a)$.

  Because there exists a set ~$F_{0}$, $F_{0}\in N(a)$,~$C\cap F_{0} =\emptyset$, so secondly we can pick~$a$. Then the remaining~$d\{x,y\}$~that we have not choose attribute satisfies does't containing~$a$, and does't contained in any attribute set in~$N(a)$. Then we can pick the attribute in the remaining~$d\{x,y\}$~but outside~$F_{0}$. Then ~$a\in B$. It contradicts with the condition of~$a$~is a unnecessary attribute. $\Box$

\begin{thero}
~$C_{D}=\{d_{A}(x,y),x,y\in U\}$~the covering-based rough sets of attribute,~$a\in A$~ is a relative necessary attribute iff ~$\{a\}$~is not an element of ~$C_{D}$,~and there exist a set~$C\subseteq A$,~$R_{C}\subseteq R_{\cup E(a)}$, and~$R_{C}\nsubseteq R_{a}$.
\end{thero}
\textbf{Proof.}
Sufficiency.~$\{a\}$~is not an element of ~$C_{D}$, then~$a$~is not a core attribute.

If there exists a set~$C\subseteq \cup E(a)$,~$R_{C}\subseteq R_{\cup E(a)}$, and~$R_{C}\nsubseteq R_{a}$, then for~$M\in C_{D},M\in E(a)$,~$C\cap M\neq \emptyset$. Let us assum~$C$~can not be reducted, because~$R_{C}\nsubseteq R_{a}$, then there is~$K\in N(a)$, satisfies~$C\cap K=\emptyset$, so we can pick~$a$, satisfies~$(\{a\}\cup C\subseteq B$, where ~$B$~is some reduct set. Then~$a$ is a relative necessary attribute.

 Necessity. If~$a\in A$ is a relative necessary attribute, then ~$\{a\}$~is not an element of~$C_{D}$,
 Then there exist a reduct $B\subseteq A$,~$a\in B$, and~$B\cap M\neq \emptyset$, where~$M\in E(a)$. Let~$C=B\cap \cup E(a)$. Then we can proof $C\subseteq \cup E(a)$, and for~$M\in C_{D}, M\in E(a)$,~$C\cap M\neq \emptyset$, then~$R_{C}\subseteq R_{\cup E(a)}$.

 If $R_{C}\subseteq R_{a}$, then for~$M\in C_{D},M\in N(a)$,~$C\cap M\neq \emptyset$, then for the ruduct~$B$, ~$a\notin B$. It contradicts with the condition of~$a$~is a relative necessary attribute. $\Box$

\begin{coro}
~$C_{D}=\{d_{A}(x,y),x,y\in U\}$~the covering-based rough sets of attribute,~$a\in A$~ is a relative necessary attribute
iff~${x}$~is not an element of ~$C_{D}$, and there exists~$C\subseteq A$, and~$F\in N(a)$,~for any set~$K\in E(a)$,~and~$C\cap K \neq \emptyset$, but~$C\cap K =\emptyset$.
\end{coro}
\textbf{Proof.} From the proof of Theorem 3.7, it holds. ~$\Box$

In the theorems above, we want to say that attribute~$a$~is unnecessary, if and only if for any attribute set~$C$, if the partition it induces is finer than~$E(a)$, then it must be finer than ~$a$. And for the core and relative necessary attribute, it doesn't hold.

\section{Attribute characters in topology}
An information system~$(U,A)$,~$|U|=n$, for~$x,y\in U$,~$d_{A}(x,y)=\{a\in A|V_{a}(x)\neq V_{a}(y)\}$.
The discernibility matrix of the information system~$(U,A)$ is~$D_{A}=\{d_{A}(x,y)|x,y\in U\}$.
So~$D_{A}$~is the power set of~$A$~whose elements are all the discernibility set.

In topology, the concept of refinement are important. It's widely studied in compact set, pre-compact set, and so on. Precise-refinement is an extension of refinement in topology.In the following, we will study the properties of attribute form Precise-refinement and refinement.
\begin{definition}
For two class of power set $A$~and~$B$~of X,~$A,B\in 2^{X}$, we call~$A$~precise-refines~$B$~iff for any ~$K_{A}\in A,$~$K_{A}\subseteq K_{B}$, for some~$K_{B}\in B$, and for any $K_{B}\in B$~there exists some~$K_{A}\in A$,
such that~$K_{A}\subseteq K_{B}$$[7]$.
\end{definition}
\begin{definition}
For two class of power set $A$~and~$B$~of $X$,~$A,B\in 2^{X}$, we call $A$~is a refinement of $B$~iff for any~$K_{B}\in B$, there exists some~$K_{A}\in A$, such that~$K_{A}\subseteq K_{B}$.
\end{definition}
\begin{thero}
For the discernibility matrix~$D_{A}$~of the information system~$(U,A)$,
~$a\in A$~is a unnecessary attribute iff there exists ~$M\subseteq D_{A}$,~$M\cap N(a)=\emptyset$, and~$M$~ precise-refines~$N(a)$.
\end{thero}
\textbf{Proof.} Sufficiency. If there exists~$M\subseteq D_{A}$,~$M\cap N(a)=\emptyset$, and ~$M$~precise-refines ~$N(a)$, then for any reduct set $B$, and for~$\forall d\{x,y\}\in M$,~$B\cap d\{x,y\}\neq\emptyset$, because ~$M$~ refines $N(a)$,
for~$\forall K\in N(a)$, there exists~$d\{x,y\}\in M$, satisfies~$d\{x,y\}\subseteq K$, then~$(B-\{a\})\cap K\neq\emptyset$, then~$a\notin B$.

Necessity. If there doesn't exist~$M\subseteq D_{A}$,which satisfies~$M\cap N(a)=\emptyset$, and $M$~precise-refines~$N(a)$,
then there exist some ~$K\in N(a)$, satisfies~$d\{x,y\}\nsubseteq K$~where~$d\{x,y\}\in D_{A}$, and
~$a\notin d\{x,y\}$.
So for any $d\{x,y\}\in D_{A}$,~$a\notin d\{x,y\}$,~$d\{x,y\}-K\neq \emptyset$. Let we assume~$K=d\{x',y'\}$, for some ~$x',y'\in U$. Because~$K\cap(A-K)=\emptyset$, so~$(x',y')\in IND(A-K)$,
so we can proof~$\{a\}\cap RED(A-K)$ is a reduct of~$A$.~$\Box$
\begin{coro}
For the discernibility matrix~$D_{A}$~of the information system~$(U,A)$, $a\in A$ is a unnecessary attribute iff there exists~$M\subseteq E(a)$, and~$M$~precise-refines~$N(a)$.
\end{coro}
\textbf{Proof.} From Theorem 4.1, if $M\subseteq D_{A}$,~$M\cap N(a)=\emptyset$, and~$M$~precise-refines~$N(a)$, then ~$M\subseteq E(a)$, then it's easy to proof. $\Box$
\begin{thero}
For the discernibility matrix $D_{A}$~of the information system~$(U,A)$,~$a\in A$~is a unnecessary attribute iff ~$E(a)$~is a refinement of~$N(a)$.
\end{thero}
\textbf{Proof.} It holds from the proof of Corollary 4.1.~$\Box$.
\begin{thero}
For the discernibility matrix ~$D_{A}$~of the information system~$(U,A)$, $a\in A$~ is a relative necessary attribute iff~$a\notin D(A)$, and there doen't exist a set $M$,which satisfies~$M\subseteq E(a)$, and $M$~precise-refines $N(a)$.
\end{thero}
\textbf{Proof.} It's obvious from Corollary 4.2. $\Box$
\begin{thero}
For the discernibility matrix~$D_{A}$~of the information system ~$(U,A)$, ~$a\in A$~ is a relative necessary attribute iff~$E(a)$~is not finer than~$N(a)$, and~$\{a\}\notin D_{A}$.
\end{thero}
\begin{coro}
For the discernibility matrix~$D_{A}$~of the information system~$(U,A)$,~$a\in A$~is a relative necessary attribute iff~$\{a\}\notin D_{A}$, and for any~$a\notin d\{x,y\}\in D_{A}$, there  exists  some~$K\in N(a)$, satisfies ~$d\{x,y\}\nsubseteq K$.
\end{coro}
\textbf{Proof.} It can be concluded by the proof of Theorem 4.1. $\Box$

\begin{exam}
An information system~$(U,A)$, the domain set~$U=\{1,2,3,4,5\}$,\\the attribute set
$A=\{a_{1},a_{2},a_{3},a_{4}\}$. For simplicity, we give the partitions by every attribute directly.\\

$\hspace{1cm}$$S_{a_{1}}=\{\{1,2\},\{3,4\},\{5\}\}$,

$\hspace{1cm}$$S_{a_{2}}=\{\{1,2,3\},\{4,5\}\}$,

$\hspace{1cm}$$S_{a_{3}}=\{\{1,2,4\},\{3,5\}\}$,

$\hspace{1cm}$$S_{a_{4}}=\{\{1,2,3,4\},\{5\}\}$.\\

The discernibility matrix is:

\begin{tabular}{cccccc}
\hline
   $U\times U$ & 1 & 2 & 3 & 4 & 5\\
\hline

1 & $\emptyset$ & $\emptyset$ & $\{a_{1},a_{3}\}$ & $\{a_{1},a_{2}\}$ & $\{a_{1},a_{2},a_{3},a_{4}\}$\\

2 & \  & $\emptyset$ & $\{a_{1},a_{3}\}$ & $\{a_{1},a_{2}\}$  & $\{a_{1},a_{2},a_{3},a_{4}\}$\\

3 & \ & \ & $\emptyset$ & $\{a_{2},a_{3}\}$ & $\{a_{1},a_{2},a_{4}\}$\\

4 & \ & \ & \ & $\emptyset$ & $\{a_{1},a_{3},a_{4}\}$\\

5 & \ & \ & \ & \ & $\emptyset$\\
 \hline

\end{tabular}

\   $\hspace{15cm}$\\
All the reducts are ~$\{a_{1},a_{2}\}, \{a_{1},a_{3}\}, \{a_{2},a_{3}\}$. $a_{1},a_{2},a_{3}$~are the relative attributes,~$a_{4}$~is the unnecessary attribute,and there are no core attributes.

Let we see the relation between ~$E(a)$~and~$N(a)$~for every attribute~$a\in A$.

(1)~$N(a_{1})=\{a_{1},a_{2}\},\{a_{1},a_{3}\},\{a_{1},a_{2},a_{4}\},\{a_{1},a_{3},a_{4}\},\{a_{1},a_{2},a_{3},a_{4}\}$,

~~~~$E(a_{1})=\{a_{2},a_{3}\}$.

 Because~$\{a_{1}\}\notin N(a)$,~$E(a_{1})$~is a refinement of~$N(a_{1})$, or there doesn't exist subset~$M\subseteq E(a_{1})$, such that~$M$~precise-refines $N(a_{1})$, then~$a_{1}$~is a relative necessary attribute.

(2)The situation of~$a_{2},a_{3}$~is just the same as~$a_{1}$.

(3)~$N(a_{4})=\{\{a_{1},a_{2},a_{4}\},\{a_{1},a_{3},a_{4}\},\{a_{1},a_{2},a_{3},a_{4}\}$,

~~~~$E(a_{4})=\{a_{1},a_{2}\}, \{a_{1},a_{3}\}, \{a_{2},a_{3}\}$~.

Because~$E(a_{4})$~is a refinement of~$N(a_{4})$, or $E(a_{4})$~precise-refines~$N(a_{4})$, then~$a_{4}$~is a unnecessary attribute.

\end{exam}
In the following, we give one more convenient method to judge the feature of attributes.

For the discernibility matrix~$D_{A}$~of the information system~$(U,A)$, we call~$d\{x,y\}\in D_{A}$~ a reducible element if there exist another~$d\{x',y'\}\in D_{A}$, satisfies~$d\{x',y'\}\subset d\{x,y\}$.

We can delete all the reducible elements, and all the reduct of ~$A$~ remain unchanged. So we denote ~$D_{reducible}$~as the subset of~$D_{A}$~whose elements are all the reducible elements of~$D_{A}$, and denote~$D_{reduct}=D_{A}-D_{reducible}$.
\begin{thero}
For the discernibility matrix ~$D_{A}$~of the information system ~$(U,A)$, $a\in A$,~$a$~is a relative necessary attribute iff ~$a\in (\cup D_{reduct})$,and ~$\{a\}\notin D_{reduct}$.
\end{thero}
\textbf{Proof.} Sufficiency. For ~$K\in D_{A}$,~if there exists another ~$K'\subset K$, then for any reduct ~$B$,~$B\cap K'\neq\emptyset$, then~$B\cap K\neq\emptyset$, so we can delete~$K$~when constructing any reduct.

   For any~$a\in (\cup D_{reduct})$, then ~$a\in K$, for some~$K\in D_{reduct}$, then for any other~$K'\in D_{reduct}$, satisfies~$K'\nsubseteq K$. Then we can proof~$\{a\}\cap RED(A-K)$ is a reduct of ~$A$.
   Besides ~$a$ is not core, then ~$a$~ is a relative necessary attribute.

Necessity. If ~$a\notin \cup D_{reduct}$, then ~$N(a)\cap D_{reduct}=\emptyset$, then there exists
~$M\subseteq D_{A}$, And~$M\cap N(a)=\emptyset$,~satisfies~$M$~a refinement of~$N(a)$, then~$a$~is unnecessary.
Then from Therom4.1, it holds.
~$\Box$

\begin{thero}
For the discernibility matrix~$D_{A}$~of the information system~$(U,A)$,~$a\in A$~is a unnecessary attribute iff $a\in (A-\cup D_{reduct})$.
\end{thero}
\textbf{Proof.} It holds from Theorem 4.5. ~$\Box$

\begin{thero}
For the discernibility matrix~$D_{A}$~of the information system ~$(U,A)$,
for any~$d\{x,y\}\in D_{reduct}$, and~$\forall a \in d\{x,y\}, (d\{x,y\}-\{a\})\neq\emptyset$, there always exists a reduct~$B\subseteq A$, satisfies~$\{d\{x,y\}-\{a\}\}\cap B=\emptyset$.
\end{thero}
\textbf{Proof.} It holds from the proof of Theorem 4.5. ~$\Box$
\begin{exam}
In Example 4.1, because ~$\{a_{1},a_{2}\}\subset\{a_{1},a_{2},a_{4}\}$,~$\{a_{1},a_{3}\}\subset\{a_{1},a_{3},a_{4}\}$,~~$\{a_{1},a_{3}\}\subset\{a_{1},a_{2},a_{3},a_{4}\}$,
otherwise there are not proper subsets  for  ~$\{a_{1},a_{2}\}, \{a_{1},a_{3}\}, \{a_{2},a_{3}\}$~ in ~$D_{A}$.

Then ~$D_{reducible}=\{a_{1},a_{2},a_{4}\},\{a_{1},a_{3},a_{4}\},\{a_{1},a_{2},a_{3},a_{4}\}$,
and ~$D_{reduct}=\{a_{1},a_{2}\},\{a_{1},a_{3}\},\{a_{2},a_{3}\}$.

Then ~$\cup D_{reduct}=\{a_{1},a_{2},a_{3}\}$,~$A-\cup D_{reduct}=\{a_{4}\}$.

Because all the unnecessary attributes belong to ~$A-\cup D_{reduct}$~,all the relative  and the core attributes belong to ~$\cup D_{reduct}$~,otherwise, ~$a_{1}\notin D_{A},a_{2}\notin D_{A},a_{3}\notin D_{A}$, then we can conclude the unnecessary attribute is ~$a_{4}$,~the relative attributes are ~$a_{1},a_{2},a_{3}$.
\end{exam}
\section{The relationship among attributes}
Attribute reduct is based on the relationship between attributes, in essence. In this section, we will give the relationship between attributes and the attribute set. That may be useful for constructing a reduct.

Let us give some kinds of marks for the relationship of two attributes in the following statements, and that is valid for two attribute sets.
\begin{definition}
In An information system~$(U,A)$, for two attributes~$a,b$, we denote $A\succcurlyeq B$, iff ~$R_{A}\subseteq R_{B}$, and that means the partition that generated by~$A$~is finer than by ~$B$. So we denote $A\approx B$, if ~$R_{A}=R_{B}$, and that means they generate the same partition of~$U$.
\end{definition}
\begin{definition}
In An information system~$(U,A)$, for two attributes~$a,b$, let we denote~$a\bowtie b$, iff  for every reduct~$B$,~$a\in B\Leftrightarrow b\in B$.
\end{definition}

\begin{definition}
In An information system~$(U,A)$, for attributes~$a$~and attribute set~$C$, we denote $C\rhd a$, iff  for any reduct~$B$,~$C\subseteq B\Rightarrow a\notin B$. That means if we pick~$C$~when constructing a reduct, then~$a$~needn't be considered.
\end{definition}
So from the definitions above, we can say that the propositions of~$a\bowtie b$~ with~$A\rhd B$, or~$B\rhd A$~can't hold simultaneously.

\begin{thero}
In An information system ~$(U,A)$, for two attributes $a,b$, $A\succcurlyeq B$, iff~$\forall K\in N(b)$,~$a\in K$, and iff~$a\in Neibor(b)$. $A\approx B$ iff~$N(a)=N(b)$, and iff ~$a\in Neibor(b)$,~$b\in Neibor(a)$.
\end{thero}
\textbf{Proof.}~$a\succcurlyeq b$ $\Leftrightarrow$ ~$R_{a}\subseteq R_{b}$~$\Leftrightarrow$
~$\forall K\in N(b)$,then~$a\in K$.
$\Box$

\begin{thero}
In An information system ~$(U,A)$, for two attributes~$a,b$, $a\bowtie b$, iff
for any~$C\subseteq A$,~$R_{C}\subseteq R_{b}$(~$R_{C}\subseteq R_{a}$),
then~$R_{C}\subseteq R_{a}$(~$R_{C}\subseteq R_{b}$).
\end{thero}
\textbf{Proof.}
Sufficiency.
If there exists a reduct ~$B$,satisfies~$a\in B$,and~$b\notin B$, then ~$R_{B}\subseteq R_{b}$,
and ~$R_{B}\nsubseteq R_{a}$.

Necessity. Assume there exists a set~$C$,~$C\subseteq A$,and~$R_{C}\subseteq R_{b}$,
but~$R_{C}\nsubseteq R_{a}$.  If there is a reduct~$B$,satisfies ~$a\in B$,then we can pick some ~$C$,~$C\subseteq A$,and ~$R_{C}\subseteq R_{b}$,~$R_{C}\nsubseteq R_{a}$,and ~$C\cup\{a\}$~belongs to some reduct.~$\Box$

\begin{thero}
In An information system~$(U,A)$, for two attributes~$a,b$, $a\bowtie b$, iff
for~$C\subseteq A$,and~$R_{(C\cup\{a\})}\subseteq R_{b}$~(~$R_{(C\cup\{b\})}\subseteq R_{a}$~), then~$R_{C}\subseteq R_{a}$~(~$R_{C}\subseteq R_{b}$).
\end{thero}
\textbf{Proof.}
Sufficiency. If there exists a reduct~$B$, satisfies~$a\in B$, and~$b\notin B$, then~$R_{B}\subseteq R_{b}$, and ~$R_{B}\nsubseteq R_{a}$. Let~$C=B-\{a\}$, then~$R_{(C\cup\{a\})}\subseteq R_{b}$, but~$R_{C}\nsubseteq R_{a}$.

Necessity. Assume there exists a set~$C$,~$C\subseteq A$, and~$R_{(C\cup\{a\})}\subseteq R_{b}$,
but~$R_{C}\nsubseteq R_{a}$. If there is a reduct~$B$, satisfies ~$a\in B$, then we can pick some~$C$,~$C\subseteq A$,and ~$R_{(C\cup\{a\})}\subseteq R_{b}$,~$R_{C}\nsubseteq R_{a}$, and ~$C\cup\{a\}$~belongs to some reduct. ~$\Box$

Intuitively speaking, Theorem 4.4 means if for a reduct~$B$,~$a\in B\Rightarrow b\in B$, then for any set~$C$, if the partition that generated by~$C\cup\{a\}$~is finer than~$b$, then $C\cup\{a\}$~can't belongs to some reduct.
\begin{coro}
In An information system~$(U,A)$, for two attributes~$a,b$, $a\bowtie b$, iff
for~$\forall K\in(N(b)-N(a))$~(~$\forall K\in(N(a)-N(b))$),~$C\subseteq A$,and ~$C\cap K\neq\emptyset$, then for~$\forall K\in(N(a)$~($\forall K\in(N(b)$)~$C\cap K\neq\emptyset$.
\end{coro}
\textbf{Proof.} For ~$\forall K\in(N(b)-N(a))$,~$C\cap K\neq\emptyset$~$\Leftrightarrow$~$R_{(C\cup\{a\})}\subseteq R_{b}$.

For~$\forall K\in N(a)$~,~$C\cap K\neq\emptyset$$\Leftrightarrow$~$R_{C}\subseteq R_{a}$.

Then from Theorem 5.3, it concludes.~$\Box$\\

For a attribute set~$C$~and an attribute~$a$, we have~$C\preceq a\Rightarrow C\rhd a$, but
~$C\rhd a\nRightarrow C\preceq a$. In the following, an example is given to illustrate this.
\begin{exam}
\ \ Let $U=\{1,2,3,4,5\}$,
$\mathbf{A}=\{a_{1},a_{2},a_{3}\}$ is a family of equivalence
relations, and the partitions they generate are as follows.

$\hspace{1cm}$$R_{a_{1}}=\{\{1,2,3\},\{4,5\}\}$,

$\hspace{1cm}$$R_{a_{2}}=\{\{1,2\},\{3,4,5\}\}$,

$\hspace{1cm}$$R_{a_{3}}=\{\{1,3\},\{2,4,5\}\}$.

 The discernibility matrix is:

\

\begin{tabular}{cccccc}
  $U\times U$ & 1 & 2 & 3 & 4 & 5 \\
  \hline
1 & $\emptyset$ & $\{a_{3}\}$ & $\{a_{2},a_{3}\}$ & $\{a_{1},a_{2},a_{3}\}$ & $\{a_{2},a_{3}\}$\\

2 & \  & $\emptyset$ & $\{a_{2},a_{3}\}$ & $\{a_{1},a_{2}\}$  & $\{a_{1},a_{2}\}$\\

3 & \ & \ & $\emptyset$ & $\{a_{1},a_{3}\}$ & $\{a_{1},a_{3}\}$\\

4 & \ & \ & \ & $\emptyset$ & $\emptyset$\\

5 & \ & \ & \ & \ & $\emptyset$

  \end{tabular}
  \\Then ~$N(a_{1})=\{\{a_{1},a_{2}\},\{a_{1},a_{3}\},\{a_{1},a_{2},a_{3}\}\}$,
  ~$E(a_{1})=\{\{a_{3}\},\{a_{2},a_{3}\}\}$.

  All the  reducts are~$\{a_{1},a_{3}\}$~and~$\{a_{2},a_{3}\}$, so for any reduct~$B$,\\
if ~$a_{2}\in B$, then~$a_{1}\notin B$, so let ~$C=\{a_{2}\}$, then~$C\rhd a_{1}$, but the partition that generated by ~$C$~is not finer than ~$a_{1}$.
\end{exam}

The next we give an equivalent conditions for~$C\rhd a$.
\begin{thero}
In An information system~$(U,A,V)$, for  attribute~$a$, and attribute set~$C$, $C\rhd a$, iff
 any attribute set~$D$, for~$\forall K\in \{E(a)-N(C)\}$, satisfies~$D\cap K\neq \emptyset$,then~$R_{C\cup D}\subseteq R_{a}$,where~$N(C)=\{d(x,y)|d(x,y)\cap C\neq\emptyset,d(x,y)\in D_{A}\}$.
\end{thero}
\textbf{Proof.}
Sufficiency.
 For any reduct~$B$, then for~$\forall K\in \{E(a)-N(C)\}$, satisfies~$B\cap K\neq \emptyset$. If~$C\subseteq B$, then for~$\forall K\in E(a)$, satisfies~$B\cap K\neq \emptyset$, then~$R_{B}\subseteq R_{a}$, then~$a\notin B$.
 Necessity.
 If there exists an attribute set~$D$, for~$\forall K\in \{E(a)-N(C)\}$, satisfies~$D\cap K\neq \emptyset$, then for ~$\forall K\in E(a)$, satisfies~$(C\cup D)\cap K\neq \emptyset$, but~$R_{C\cup D}\nsubseteq R_{a}$, that means to say there exists a ~$K_{0}\in N(a)$~,~$(C\cup D)\cap K_{0}= \emptyset$.

 If~$C$~is contained in some reduct, then by Theorem 4.5, we can construct a reduct~$B$,
 satisfies ~$(C\cup \{a\}\cup RED(D))\subseteq B$, then ~$a\in B$.It contradicts with the condition of $C\rhd a$. ~$\Box$
\section{A kind of attribute reduct algorithm based on $E(a)$}
In [3], Yao proposed a row-wise simplification reduct construction algorithm. His thought is described as follows:

It is based on the discernibility matrix~$M$~whose element in~$i-th$~row,~$j-th$~column is~$M(i,j)$.
For an non-empty~$M(x,y)$, it 's simplified into three steps. First, absorb~$M(i,j)$~by all the elements in
~$M$,so there are not proper subset of~$M(i,j)$~in~$M$. Secondly, select an attribute~$a$~in~$M(i,j)$~for constructing a reduct, and absorb elements in~$M$~by ~$\{a\}$.Thirdly,for all ~$M(x,y)\neq\emptyset,M(x,y)\in M$, let ~$M(x,y)=M(x,y)-M(i,j)$. Then back to the first step to continue. In the end, a reduct is constructed.\\
The row-wise simplification reduct construction algorithm proposed by Yao in [3] is recalled in the following:\\
\textbf{Input:}The discernibility matrix~$M$~of an information table~$S$.\\
\textbf{Output: }A reduct ~$R$.\\
\textbf{for} ~$i=2$~to ~$n$ \textbf{do} \{\\
\textbf{for }~$j=1$~to ~$i-1$\{\\
if ~$M(i,j)\neq\emptyset$\{\\
//Absorb~$M(i,j)$~by every non-empty element in ~$M$\\
for every non-empty element element ~$M(x',y')\in M$~\textbf{do}\\
if  ~$M(x',y')\subseteq M(i,j)$~then\\
~$M(i,j)=M(x',y')$.\\

//Divide ~$M(i,j)$~into two parts\\
\textbf{select} an attribute ~$a$~from ~$M(i,j)$;\
~$A=M(i,j)-\{a\}$;\\
~$M(i,j)=\{a\};$\\

//simplify every non-empty element in~$M$\\
\textbf{for} every non-empty element ~$M(x',y')\in M$~do\\
\textbf{if} ~$a\in M(x',y')$~then\\
~$M(x',y')=\{a\}$~;\\
\textbf{else}\\
~$M(x',y')=M(x',y')-M(i,j)$~;\\
\}// end if\\
\}// end for loop of ~$j$\\
\}// end for loop of ~$i$\\

In the section above, we point out that~$E(a)$~can decide the attribute~$a$~whether or not belongs to some reduct, so we construct a algorithm with the use of~$E(a)$~and the row-wise simplification reduct construction algorithm. The detailed algorithm is given in the following.

In the following,we denote a reduct set~$RED_{E(a)}$~with respect to~$E(a)$~iff ~$RED_{E(a)}\subseteq \cup E(a)$, and for any~$K\in E(a)$,~$RED_{E(a)}\cap K\neq\emptyset$, and for any proper subset of~$RED_{E(a)}$, that's not true.

\textbf{A algorithm with respect to~$E(a)$:}

\textbf{Input:} The set~$C_{D}$~of subsets of attributes transformed from the discernibility matrix~$D_{A}$;
   //Let~$C_{D}=\{C(i);1\leqslant i\leqslant N\}$, where~$N$~represents the number of nonempty-set elements in  the discernibility matrix~$D_{A}$; so~$ N\leqslant (n^{2}-n)/2$,~$n$~is the cardinal of the object set~$U$.\\
\textbf{Output:} A reduct R.\\
first let ~$R=\emptyset$\\
while (~$C\neq \emptyset$~)

\noindent\textbf{first step:}\\
select an attribute ~$a$,~$a\in C(1)$ ;\\
compute ~$N(a)$,~$E(a)$;\\
\textbf{second step:}\\
compute a reduct respect to~$E(a)$~with the row-wise simplification reduct construction algorithm in [3]; we denote the reduct with respect to~$E(a)$~as~$RED_{E(a)}$;\\
let ~$R=R\cup RED_{E(a)}$;\\
\textbf{third step:}\\
\textbf{if} there exists a~$K\in N(a)$,such that~$RED_{E(a)}\cap K=\emptyset$;\\
then let ~$R=R\cap \{a\}$; \\
\textbf{else continue};\\
\textbf{forth step:}\\
 \textbf{for} every ~$C(i')\in C$, let~$C(i')=C(i')-\cup N(a)$;
 //because ~$E(a)\subseteq \cup N(a)$, thus ~$C(i')-\cup N(a)=C(i')-\cup N(a)-\cup E(a)$ \\
 \textbf{if} ~$C=\emptyset$;\\
 \{\textbf{Output} ~$R$;     //~$R$~is a reduct\\
 \textbf{End.}\}\\
 \textbf{else return};  //goto the first step\\
\}\\
\}\\
In the second step, we can also return to the first step to compute a reduct with respect to~$E(a)$, and so the absorption operation needn't be used. We don't give more description because of the limitation of length.

\begin{thero}
The algorithm respect to~$E(a)$~can construct a reduct of~$A$.
\end{thero}
\textbf{Proof.} In the second step of the algorithm, we can pick  a reduct respect to~$E(a)$.

After the third step, we can pick a reduct with respect to ~$E(a)\cup N(a)$.

In the forth step,for~$C(i')\in C$, if ~$C(i')\notin E(a)\cup N(a)$, then~$C(i')-\cup N(a)\neq\emptyset$.

From the method of induction, in the following of the operations of the algorithm, we will construct a reduct respect to~$C-E(a)-N(a)$.

 Because after the forth step, for any~$C(i')\in C$,~$C(i')\neq\emptyset$, and for any~$K\in N(a)$, or~$K\in  E(a)$, such that $C(i')\cap K=\emptyset$,
 thus the union of the reduct with respect to~$E(a)\cup N(a)$~and the reduct with respect to~$A-\cup E(a)-\cup N(a)$~is also a reduct.

 Then it's easy to proof that's a reduct of~$A$.~$\Box$\\

In [3], Yao used an example to inspect the row-wise simplification reduct construction algorithm, now we will quote this example in [3] to exam the algorithm with respect to~$E(a)$.
\begin{exam}

 The discernibility matrix is:
\

\begin{tabular}{ccccccc}
  $U\times U$ & 1 & 2 & 3 & 4 & 5& 6 \\
  \hline
1 & $\emptyset$ & $\emptyset$ & $\emptyset$ & $\{a,b,f\}$ & $\{a,c\}$ & $\{a,d\}$\\

2 & \  & $\emptyset$ & $\emptyset$ & $\{c,d,f\}$  & $\{b,d\}$ & $\{b,c\}$\\

3 & \ & \ & $\emptyset$ & $\{b,e,f\}$ & $\{c,e\}$ & $\{d,e\}$\\

4 & \ & \ & \ & $\emptyset$ & $\emptyset$ & $\emptyset$\\

5 & \ & \ & \ & \ & $\emptyset$ \ & $\emptyset$\\

6 & \ & \ & \ & \ & \ & $\emptyset$
  \end{tabular}

 The steps that we use the algorithm are as follows:

 The set~$C_{D}$~of subsets of attributes transformed from the discernibility matrix ~$D_{A}$~is\\
 ~$C_{D}=\{\{a,b,f\},\{a,c\},\{a,d\},\{c,d,f\},\{b,d\},\{b,c\},\{b,e,f\},\{c,e\},\{d,e\}\}$.\\

 \noindent\textbf{first step:}\\
select the attribute ~$a$,~$a\in C(1)$ ;\\
compute ~$E(a)=\{\{c,d,f\},\{b,d\},\{b,c\}\}$,~$N(a)=\{\{a,b,f\},\{a,c\},\{a,d\}\}$.\\

\textbf{second step:}\\
compute a reduct respect to~$E(a)$~with the row-wise simplification reduct construction algorithm in[3];
then~$RED_{E(a)}=\{c,b\}$.\\
Then ~$R=RED_{E(a)}$.\\

\textbf{third step:}\\
because for~$\{a,d\}\in N(a)$,such that~$RED_{E(a)}\cap \{a,d\}=\emptyset$,\\
then ~$R=RED_{E(a)}\cap \{a\}=\{a,b,c\}$; \\
\textbf{forth step:}\\
for every ~$C(i')\in C$~, let ~$C(i')=C(i')-\cup N(a)$.\\
Then  ~$C_{D}=\{\{e\}\}$.\\
Then  ~$R=RED_{E(a)}\cup \{e\}$; then ~$R=\{a,b,c,e\}$.\\
\end{exam}
\textbf{Complexity analysis of algorithms:}

All the elements need the absorption operation in he row-wise simplification reduct construction algorithm, and the time that each absorption operation cost is about~$0(n^{2})$, so that's a complex operation. In the algorithm based on~$E(a)$, the absorption operation is performed only in~$E(a)$, but not in all the elements of the discernibility matrix. Otherwise, in the forth step,we let~$C(i')=C(i')-\cup N(a)$,for every~$C(i')\in C$~, so we can delete more attributes in the matrix, and remain the elements of the discernibility matrix with fewer attributes, So that may be more simple when constructing a reduct.

The thought of using~$E(a)$~in this algorithm is  useful for  attribute reduct. Because when we construct the reduct of~$E(a)$, then~$a$~must or can't select is determined. In the next, we can delete the union of~$E(a)$~,then the attribute set with fewer elements remained that is more simple.

Let we study the complexity of  the row-wise simplification reduct construction algorithm for Example 6.1 which is first used by Yao in [3].The algorithm traverses the discernibility matrix~$D_{A}$~three times, and  every time, the algorithm experienced the operations which contain absorption,partition for  some attribute set,and inspection. So the total of  times for the algorithm which used in Example 6.1 is about~$36*3*3$, where~$36$~is the number of the elements in the discernibility matrix~$D_{A}$.

In the algorithm based on~$E(a)$, the operations we experienced included computing~$E(a)$~and~$N(a)$,
computing a reduct respect to~$E(a)$~with the row-wise simplification reduct construction algorithm, comparing
~$RED_{E(a)}$~with ~$N(a)$, and for every ~$C(i')\in C$, computing~$C(i')=C(i')-\cup N(a)$. So the total of  times for this algorithm is about~$9*4$~where~$9$~is the number of elements in~$C_{D}$~in Example 6.1.

In the algorithm based on~$E(a)$, we only use the row-wise simplification reduct construction algorithm for~$E(a)$, and after we pick the reduct of~$E(a)$~and~$N(a)$,we will delete all the attribute in~$\cup N(a)$, then less attributes are left. The time complexity for the worst is about~$0(n^{2}*ln^{3}n)$, So this algorithm maybe more efficient when constructing a reduct for some cases.

\section{Conclusion}
Attribute reduct is a NP-hard problem, many methods are proposed to to find all the reducts or a single reduct, but there haven't the optimal solutions for it, so it's still a problem worth of intensive researching. In some degree, constructing a reduct is based on the relationship between attributes. In this paper, we give the substantive conclusions of attribute character in terms of the relationship among attributes. We point out that the attribute is whether unnecessary or relative necessary is determined by the relationship between $E(a)$ and $N(a)$ in virtual, and whether unnecessary or relative necessary with respect to some reduct is determined by the relationship of this reduct, $E(a)$ and $N(a)$.

Furthermore, we give some easy and clear description of the attribute features in topology. Besides, some relationships between attributes are given and that is helpful for constructing reducts. Finally, a method of attribute reduct is given based on $E(a)$. This kind of thought is maybe useful for the further study of attribute reduct in rough set theory.

\section*{Acknowledgement}

The authors would like to thank the anonymous reviews for their
constructive comments. This work is supported by grants from
National Natural Science Foundation of China under Grant (Nos.
71140004, 10971186). \vskip  6mm

\section*{References}
\noindent$[1]$ Z. Pawlak, Rough sets, International Journal of Computer and Information Sciences 11 (1982) 341-356.\\
$[2]$ Z. Pawlak, Rough sets: theoretical aspects of reasoning about data, Kluwer Academic Publishers, Dordrecht, MA, 1991.\\
$[3]$ Y. Yao, Y. Zhao, Discernibility matrix simplification for constructing attribute reducts, Information sciences 179 (2009) 867-882.\\
$[4]$ W. Zhang, G. Qiou, Uncertain decision making based on rough sets, Publishing Company of Qinghua University, 2005.\\
$[5]$ W. Zakowski, Approximations in the space ~$(U,\pi)$~, Demonstratio Mathematica 16 (1983) 761-769.\\
$[6]$ A. Skowron, C.Rauszer, The discernibility matrices and functions in information systems, in: R.Slowi$\acute{n}$ski(Ed.), Intelligent Decision Support, Handbook of Applications and Advances of the Rough Sets Theory, Kluwer, Dordrecht, 1992.\\
$[7]$ G. Gao, The theory of topology spaces[M], Science Publishing Company of Beijing, 2008.\\
$[8]$ T. Lin, Introduction to special issues on data mining and granular computing, International Journal of Approximate Reasoning 40 (2005) 1-2.\\
$[9]$ J. Mi, y. Leung, W. Wu, An uncertainty measure in partition-based fuzzy rough sets,International Journal of Approximate Reasoning 34 (2005) 77-90.\\
$[10]$  Yee. Leung, Manfred M. Fischer, W. Wu, J. Mi, A rough set approach for the discovery of classi?cation rules in interval-valued information systems, International Journal of Approximate Reasoning 47 (2008) 233-246.\\
$[11]$ Q. He, C. Wu, D. Chen, S. Zhao, Fuzzy rough set based attribute reduction
for information systems with fuzzy decisions, Knowledge-Based Systems 5 (2011) 689-69.\\
$[12]$ W. Wu, Attribute reduction based on evidence theory in incomplete decision systems, Information sciences 178 (2008) 1355-1371.\\
$[13]$ Y. Yao, Constructive and algebra methods of theory of rough sets, Information sciences 109 (1998) 21-47.\\
$[14]$ L. Feng, T. Li, D. Ruan, S. Gou, A vague-rough set approach for uncertain
knowledge acquisition Original Research Article, Knowledge-Based Systems 6 (2011) 837¨C843.\\
$[15]$ Y. Yao, Y. Zhao, Attribute Reduction in decision-theoretic rough set models, Information sciences 178 (2008) 3356-3373.\\
$[16]$ J. Wang, J. Wang, Reduction algorithms based on discernibility matrix: the ordered attributes method, Journal of Computer Science and Technology 16 (2001) 489-504.\\
$[17]$ K. Zhao, J. Wang, A reduction algorithm meeting users' requirements, Journal of Computer Science and Technology 17 (2002) 578-593.\\
$[18]$ Y. Yao, Y. Zhao, J. Wang, S. Han, A model of machine learning based on user preference of attributes, Proceedings of International Conference on Rough Sets and Current Trends in Computing, 2006, pp. 587-596.\\
$[19]$ W. Zhang, J. Mi, W. Wu, Approaches to knowledge reductions in inconsistent systems, International Journal of Intelligent Systems 18 (2003) 989-1000.\\
$[20]$ Z. Pawlak, A. Skowron, Rough sets and boolean reasoning, Information Sciences 177 (2007) 41-73.\\
$[21]$ D. Chen, C. Zhang, Q. Hu, A new approach to attribute reduction of consistent and inconsistent covering decision systems with covering rough sets, Information Sciences 177 (2007) 3500-3518.\\
$[22]$ D. $\acute{S}$lezak, W. Ziarko, The investigation of the Bayesian rough set model, International Journal of Approximate Reasoning 40 (2005) 81-91.\\
 $[23]$ Y. Qian, J. Liang, C. Dang, Knowledge structure, knowledge granulation and knowledge distance in a knowledge base, International Journal of Approximate Reasoning 50 (2009) 174-188.\\
$[24]$ T. Yang, Q. Li, The reduction and fusion of fuzzy covering systems based on the evidence theory, International Journal of Approximate Reasoning 53 (2012) 87-103.\\
$[25]$ M. Hall, G. Holmes, Bench marking attribute selection techniques for discrete class data mining, IEEE Transactions on Knowledge and Data Engineering 15 (2003) 1437-1447.\\
$[26]$ G. Liu, Generalized rough sets over fuzzy lattices, Information Sciences 178 (2008) 1651-1662.\\
$[27]$ L. Zadeh, Fuzzy logic$=$computing with words, IEEE Transactions on Fuzzy Systems 4 (1996) 103-111.\\
$[28]$ Z. Pei, D. Pei, Li Zheng, Topology vs generalized rough sets, International journal of Approximate Reasoning 52 (2001) 231-239.\\
$[29]$ X. Li, S. Liu, Matroidal approaches properties of rough sets via closure operators, International Journal of Approximate Reasoning 53 (2012) 513-527.\\
$[30]$ A. Salama, Some topological properties of rough sets with tools for data mining, International Journal of Computer Science 8 (2011) 588-595.\\
$[31]$ W. Zhu, Topological approaches to covering rough sets based on relations,  Information sciences 177 (2007) 1499-1508.\\
$[32]$ Y. Chen, D. Miao, R. Zhang, K. Wu, A rough set approach to feature selection based on power trees, Knowledge-based Systems 24 (2011) 275-281.\\
$[33]$ W. Zhu, F. Wang, On three types of covering rough sets, IEEE Transactions on Knowledge and Data Engineering 19 (2007) 1131-1144.\\

\end{document}